\documentclass{ifacconf}

\usepackage{graphicx}   
\usepackage{natbib}  
\usepackage{booktabs}
\usepackage{multirow} 

\usepackage{graphicx}
\usepackage{tikz}
\usepackage{xcolor}
\usepackage{endnotes}



\definecolor{kBest}{HTML}{00008B}   
\definecolor{kOne}{HTML}{D55E00}    
\definecolor{kGreen}{HTML}{009E73} 
\definecolor{kPink}{HTML}{CC79A7}  
\definecolor{kYellow}{HTML}{F0E442} 
\definecolor{kGrey}{HTML}{999999}   
\definecolor{kCyan}{HTML}{56B4E9}   
\usetikzlibrary{positioning, arrows.meta, calc, shapes.geometric, shadows}
\usepackage{graphicx}
\usepackage{epstopdf}

\usepackage{times}
\usepackage{algorithm}

\usepackage[noend]{algpseudocode}
\usepackage{float}

\theoremstyle{plain}
\usepackage{amsmath}
\usepackage{amssymb}

\newcommand{\Qstat}{Q^{st}}
\newcommand{\StSet}{\mathcal{S}}
\newcommand{\AcSet}{\mathcal{A}}
\newcommand{\TrF}{T}
\newcommand{\reward}{r}

\newcommand{\MDPi}{$\mathcal{P}_i = (Pos\times\tilde\StSet_i,\AcSet_i,\TrF_i,\reward_i,\gamma_i)$}
\DeclareMathOperator{\argmax}{argmax}
\newcommand{\AgentSet}{\mathcal{I}}
\newcommand{\ObSet}{\Omega}
\newcommand{\ObF}{\mathcal{O}}

\newcommand{\DecPOMDP}{\mathcal{DP} = (\AgentSet,\StSet,\{\AcSet_i\},\TrF,\reward,\{\ObSet_i\},\gamma)}

\newcommand{\model}{\mathcal{M}_i}

\usepackage{url}

\begin{document}
\begin{frontmatter}

\title{Decentralized Safe Multi-Agent Reinforcement Learning via Predictive Shielding}

\author[First]{Yacine El Yamani} 
\author[Second]{Hanna Krasowski} 
\author[First]{Elena Vanneaux}

\address[First]{ENSTA, IP Paris, Palaiseau, France (email: yacine.elyamani, elena.vanneaux@ensta.fr )}
\address[Second]{UC Berkeley, USA (e-mail: krasowski@berkeley.edu)}

\begin{abstract}Environments are increasingly populated by multiple robots performing independent tasks with limited prior knowledge of each other. Deploying such multi-agent systems presents significant challenges. Specifically, shifts in deployment states compared to training data can lead to poor policy performance and compromised safety. While safety shields exist to mitigate these risks, they are typically reactive, which degrades performance near unseen obstacles,and centralized, limiting their scalability. To address this, we propose a decentralized framework that integrates predictive shielding with model-based finite horizon Q-learning. This approach allows agents to safely adapt their pre-trained policies during deployment. Furthermore, to mitigate livelocks in symmetric scenarios, we introduce a communication-free protocol for conflict resolution\endnote{Source code: \tiny\url{https://github.com/YacineEY/Decentralized-Safe-Multi-Agent-Reinforcement-Learning-via-Predictive-Shielding.git}}\makeatletter\xdef\monNumeroGitHub{\the\value{endnote}}\makeatother.

\end{abstract}


\end{frontmatter}

\section{Introduction}

In recent years, it has become increasingly common for multiple robots, often based on reinforcement learning (RL) agents, with distinct functionalities to operate simultaneously within
a shared environment. For example, delivery robots transport packages and service robots perform continuous cleaning - all operating independently while sharing the same physical space \citep{Oroojlooy2022, GuoMeng}. Unlike most multi-agent RL approaches, which assume joint training and shared objectives \citep{Tang2025_DeepRL_Robotics, Oroojlooy2022}, this paper considers agents pretrained separately for individual tasks and later deployed together. Directly executing the pretrained policies may lead to collisions between agents. To ensure safety during deployment,safety shields are commonly used \citep{ChuChuFAn}.

In contrast to shields designed for joint multi-agent systems \citep{ElSayedAly,SafePOMDPOnlinePlanningviaShielding}, we assume that every robot is incorporated with an individual shield that replaces the unsafe actions proposed by the agent with a provably safe backup policy \citep{krasowski2023provably}. The individual shields are then composed together using the assume–guarantee paradigm to ensure safety for the overall multi-agent system \citep{BrorholtCompositionalShielding, Bakirtzis}. Assume-guarantee shields design does not require explicit communication between agents \citep{XiaoModelBasedDynamicShielding}, but they are commonly based on simple safety rules (e.g., stop to avoid collision). 

The composition of reactive shields with trivial backup policies can significantly degrade the task-related performance of the agents \citep{ElSayedAly}. 
To address this limitation, we propose to equip each agent with a model-based predictive shield \citep{jin2025predictive,BanerjeeDynamicMPSShield}. Unlike minimal interference shields \citep{ElSayedAly} that modify the policy only when unsafe action is about to be taken on the next step, a model predictive shield uses the environment model learned during training and local sensing data to optimize the behavior of the agent for a few steps ahead. In contrast to existing work \citep{jin2025predictive,BanerjeeDynamicMPSShield}, our shield differentiates between static and dynamic obstacles and adapts infinite-horizon RL from \citep{jin2025predictive} for static environments and finite-horizon RL for dynamic constraints \citep{BanerjeeDynamicMPSShield}. 

The proposed predictive shield aims to adapt to unseen constraints and minimize inter-agent interference through predictive optimization. However, agents may still enter a livelock, for example when symmetric actions are taken. Inspired by \cite{grover2023deadlock, ma2024deadlock}, we introduce a novel communication-free conflict resolution protocol to address this issue. The protocol randomly decides whether an agent should aggressively follow its desired path or yield to the other agent.

The main contributions of this paper are the following:
\begin{itemize}
    \item We propose a decentralized model predictive shield for non-communicating heterogeneous agents operating in shared environments. Assuming that every agent equipped with a trivial backup policy it improves the total reward of the group, defined as a sum of the rewards of every agent, while preserving the hard safety guarantees.
    \item We introduce a communication-free stochastic conflict resolution protocol for resolving deadlocks in symmetric multi-agent configurations.
     \item We demonstrate through experiments that the proposed method achieves scalability, low online computational cost, and robust performance in dense multi-agent environments.
\end{itemize}

\section{Preliminaries}
{\bf Multi-Agent Reinforcement Learning. }
A single RL agent is commonly modeled as a Markov Decision Process (MDP). Dec-POMDPs extend the traditional MDP framework to cooperative multi-agent settings, where multiple agents make decisions under uncertainty and partial observability \citep{IQL}.
\begin{defn}
    \label{Dec-PoMDP} 
   
A { Dec-POMDP} is a tuple  $$\DecPOMDP,$$
where $\AgentSet = \{1, 2, \dots, N\}$ is the set of agents, $\StSet$ is the state space, $\AcSet_i$ is the action space of agent $i$ with ${\AcSet} = \times_{i \in \AgentSet} \AcSet_i$ the joint action space, $\TrF\colon \StSet \times \AcSet \rightarrow \StSet$ is a deterministic transition function, $\reward\colon \StSet \times {\AcSet}\rightarrow \mathbb{R}$ is the joint reward function, $\Omega_i$ is the set of observations $o_{i}(s)$ available to agent $i$ with $\Omega = \times_{i \in \AgentSet} \Omega_i$ the joint observation space, and $\gamma \in [0,1)$ is the discount factor.
 $\ObF\colon \StSet \times {\AcSet} \times \Omega \rightarrow [0,1]$ is the observation function with $\ObF(s', { a}, o) = \mathbb{P}(o_{t+1} = o \mid s_{t+1} = s', {a_t} = { a})$ that gives probability of receiving observation $o$ when the system is actually in state $s$ and the joint action is ${ a}$ 
\end{defn}
The objective is to find a joint policy $\pi = (\pi_1, \dots, \pi_n)$ that maximizes the expected discounted return 
\( \mathbb{E} \left[ \sum_{t=0}^{\infty} \gamma^t r(s_t, {a_t})\right]\). Exactly solving a Dec-POMDP is NEXP-complete, even for finite horizons \citep{BernsteinPOMDP} problems. One of the approaches to find an approximate solution is IQL \citep{IQL,pmlr-v258-jin25a}, where each agent $i$ maintains its own $Q^{(m)}_i(o_i(s),a_i)$-table. All agents take actions simultaneously, using $\varepsilon$-greedy policy w.r.t. $Q^{(m)}_i(o_i(s),a_i)$, and then update $Q^{(m)}_i(o_i(s),a_i)$ according to the rule:
\begin{multline}
      \label{IQlearningUpdate}
Q_i^{(m)}(o_i(s_t), a_{i,t}) \leftarrow  Q_i^{(m)}(o_i(s_t), a_{i,t}) \\
             + \alpha\Big[r_{i,t}(o_i(s_t),a_{i,t}) +\\ \gamma \max_{a_i' \in \AcSet_i} Q_i^{(m)}(o_i(s_{t+1}), a_i') - Q_i^{(m)}(o_i(s_t), a_{i,t})\Big].
\end{multline}
The individual rewards are chosen to be such that $r(s_t,a_t) = \sum_{i\in\AgentSet} r_{i,t}(o_i(s_t),a_{i,t}).$ Independent $Q$-learning does not require coordination or communication between the agents, it scales well with respect to the number of agents, does not always converge.

{\bf Provably Safe Multi-Agent Reinforcement Learning. }
A common way to make RL controllers suitable for safety-critical applications is to pair them with post-posed safety shields~\citep{ElSayedAly}. These shields accept safe actions proposed by the agents and replace unsafe ones with a precomputed, provably safe, backup policy $\pi_{backup}$.
\begin{defn}
\label{DefSafetySpec}
A {\bf safety specification} is a sequence of state constraints that must not be violated during execution of Dec-POMDP, i.e. $s_t\in \StSet^{safe} \subseteq \StSet,\, \forall t \in \mathbb{Z}_+$.
\end{defn}

Let us assume that every agent equipped with a monitoring system $\varphi_i\colon\Omega_i\times \mathcal{A}_i \rightarrow \{0,1\}$ verifying if the action $a_i$ is safe  ($\varphi_i(o_i(s),a_i) = 1$) or unsafe ($\varphi_i(o_i(s),a_i) = 0$) in a state $s$ and a shield $\phi_i \colon \Omega_i\times\AcSet_i \rightarrow \AcSet_i$ such that for all $ a_i\in \AcSet_i$
\begin{equation}
\phi_i(o_i(s), a_i) = \left\{\begin{aligned}
&a_i& \mbox{ $\varphi_i(o_i(s),a_i) = 1$ }\\
&\pi_{backup}(o_i(s)) & \mbox{$\varphi_i(o_i(s),a_i) = 0$}\\
\end{aligned}
\right.
\end{equation}
We assume that $\varphi_i(o_i(s),\pi_{backup}(o_i(s))) = 1$ for all $s \in\StSet$.
\begin{defn}
Let us define a sequence of sets 
{\small\begin{equation}
\label{defWt}
    \mathcal{W}_t = \bigcup_{s\in \mathcal{W}_{t-1}}\bigcup_{{ a}\in\AcSet}T\big(s,\{\phi_1(o_1(s),a_1),\ldots \phi_n(o_n(s),a_n)\}\big)
\end{equation}}
 A {decentralized post-posed shield} $\phi = \{\phi_i\}, i\in\AgentSet$ ensures safety for a winning region $\mathcal{W}_0 \subseteq \StSet^{safe}$ if and only if $\mathcal{W}_t \subseteq \StSet^{safe}$, $\forall t \in \mathbb{Z}_+$.
\end{defn}
Let us remark that in Definition \ref{Dec-PoMDP} we assume the transition relation to be a deterministic function, in contrast to the more classical definition when the transition relation is the probability that action $a$ in a state $s$ at time $t$ will lead to state $s'$ at time $t+1$. We do it to ease the explanation and analysis of the proposed ideas. However, our approach could be extended to probabilistic settings, following the ideas from \citep{Bastani}.

\section{Problem Statement}

Let us consider a multi-robot system composed of $N$ agents:
\begin{equation}
\label{DefAgentSet}
    \AgentSet = \{1,\ldots,N\}.
\end{equation}

{\bf Training Phase.} Let us assume that every agent is trained independently to solve its task in the environment where no other agents are present. Every agent \( i \in \AgentSet \) is modeled as an MDP \MDPi, where $Pos$ is a set
of all possible positions of a robot in the environment, and
we suppose that this set is the same for all the agents. Hence,
every state $s_i \in S_i = Pos \times \tilde S_i$ can be decomposed into
two components $s_i = (s_{pos,i}, \tilde s_i)$, where $s_{pos,i}$ represents a
position of a robot in the environment, and $\tilde s_i$ represents other
task-related features. Note that since in this work the position space is where conflicts between agents arise, this subspace is relevant for checking safety. However, for the method of the paper it can be viewed as any space in which the conflict potentially arises. We also assume that $\StSet_i$ and $\AcSet_i$ are finite and discrete. Every agent is trained to optimize its own cumulative reward $\mathbb{E}_{\pi_i}\sum_{t=0}^\infty\gamma^tr_{i,t}(s_{i,t},a_{i,t}).$ During the training, each agent learns
\begin{enumerate}
    \item an optimal value-action function \( Q_i^{tr}\) and the corresponding optimal policy $\pi_i^{tr}(s_i) = \argmax_{a_i\in A_i} Q_i^{tr}(s_i, a_i)$.
\item a sample-based model $\model = \{(s_i,a_i,r_i,s_i')\} $ which stores the transitions of the agent during training.
\end{enumerate}

{\bf Deployment Phase.} At deployment, all $N$ agents operate simultaneously in a shared environment with joint state and action spaces
\begin{equation}
\label{defStateSet}
\StSet = \times_{i\in\AgentSet}\StSet_i, \ \AcSet = \times_{i\in\AgentSet}\AcSet_i
\end{equation}
We assume no communication between agents, so each agent treats the others as part of the environment. Agent $i$ observes its own state and partial observations of nearby agents within sensing range $\rho_i$. The observation function is
\begin{equation}
\label{defObservation}
o_i(s)={s_i}\cup_{j:|s_{pos,i}-s_{pos,j}|\leq \rho_i}o(i,j),
\end{equation}
where $o(i,j)$ always includes the position of agent $j$.

Due to partial observability and the absence of communication, deployment is a Dec-POMDP with transition function
\begin{equation}
\label{defTransitionRelation}
\TrF(s,a)=(T_1(s_1,a_1),\ldots,T_n(s_n,a_n)).
\end{equation}
We assume the transition functions $T_i$ are unchanged between training and deployment.

{\bf Objectives}. We aim to ensure safe and performant agent behavior during deployment.

{\bf \it Safety}. For any time $t\in\mathbb{Z}_+$ a position $s_{pos,t} \in Pos$ is unsafe if there is an obstacle there or there exist $i\neq j$ such that $s_{pos,i,t} = s_{pos,j,t} = s_{pos,t}$. A state $s_{i,t}$ of an agent $i$ is unsafe if $s_{pos,i,t}$ or $\tilde s_{i,t}$ are unsafe. The joint multi-agent state $ {s_t} = (s_{1,t},\ldots s_{n,t})$ is safe if all agent states are safe. We therefore define a safe set $\StSet^{safe}\subseteq\StSet$ that agents must remain in during execution

A state that is safe during training may become unsafe during deployment. We assume a decentralized communication-free safety shield $\phi={\phi_i}_{i\in\AgentSet}$ for the Dec-POMDP \eqref{DefAgentSet}-\eqref{defTransitionRelation}; its design is beyond the scope of this paper, though Section~\ref{sec:experiments} presents a suitable shield for our benchmarks.

{\bf \it Performance}. Let us assume that the agents' tasks remain unchanged, then reward functions $r_i(s_i,a_i)$ one used for training are relevant at the deployment phase as well. While the shield $\phi$ will guarantee safety during the deployment, the shielded policy $\{\pi_1^{tr},\ldots,\pi_N^{tr}\}$ not necessarily maximizes \eqref{mainCumReward}.
In this paper, we search for joint policies $\pi = \{\pi_1,\ldots,\pi_n\}$, that improve the performance of the system in terms of the cumulative reward of all agents:
\begin{equation}
\label{mainCumReward}
\mathbb{E}_\pi\left[\sum_{t=0}^{\infty} \gamma^t  \sum_{i = 1}^N r_i(s_{i,t}, \phi_i(o_i(s_t), a_{i,t}))\right].
\end{equation}

\section{Model Predictive Shield}

At any time $t$, agent $i$ can use its local observation $o_i(s_t)$, and the sampling model $M_i(s_{i,t},a_{i,t},r_{i,t},s_{i,t}')$ to locally re-plan its trajectory. 
Since every agent has a limited observability range $\rho_i$, we limit the re-planning to this range. 
The objective is to improve the contribution of the agent $i$ to the joint reward \eqref{mainCumReward}, by simulating the two-player game, where the agent $i$ tries to maximize the reward
\begin{equation}
\label{maincumrewardperagent}
\mathbb{E}_{\pi_i}\left[\sum_{t=0}^{\infty} \gamma^t r_i(s_{i,t}, \phi_i(o_i(s_t), a_{i,t}))\right],
\end{equation}
while the safety shield acts as an adversarial agent. When the shield is used in the simulation, the agent receives a negative reward $r_{sh}$, and the action is replaced with $\pi_{backup}$. 
We assume that agents can distinguish between static and dynamic safety constraints. That enables us to solve the simulated game differently for stationary versus non-stationary environmental changes. When the agent's position $s_{pos,i,t}$ leaves the region $B_{\rho_i}(s_{pos,i,0}) =\{s_{pos}\in Pos \mid  \|s_{pos} - s_{pos,i,0}\|\leq \rho_i\}$ the simulated process reaches the terminal state.
$\max_{a_i\in\AcSet_i} Q_i^{tr}(s_{i,t+1},a_i).$
Finally, the agent then takes the action that is optimal with respect to the simulated game. 
In the following, we detail how we solve the two-player game for static and dynamic constraints and explain the overall algorithm for safe deployment of independently learned Q-functions.

\subsection{Predictive shield  for static safety constraints }

To handle static safety constraints, we apply the methodology presented in \citep{jin2025predictive}.
Let us remark that agent $i$ treats the others as dynamical obstacles, hence at this stage we can know that there are no other agents in our sensing range, consequently, in \eqref{maincumrewardperagent}, $o_i(s_t)$ is equal to $s_{i,t}$.
For every agent $i$, let us define a sequence of reachable sets
\begin{equation}
\label{reachSets}
\begin{split}
    R_i(0 \mid t, s_{i,t}) &= s_{i,t}, \\ 
    R_i(k \mid t, s_{i,t}) 
    &= \bigcup_{a_i \in \AcSet_i}
       \bigcup_{s_i \in R_i(k-1 \mid t, s_{i,t})}
       T_i(s_i, a_i).
\end{split}
\end{equation}
We then introduce the planning horizon $k_{stat}$ and simulate the game that optimizes \eqref{maincumrewardperagent} in the local region $R_{i, k_{stat}}.$ Every state $s_{i,t}$ beyond $R_{i, k_{stat}}$ is terminal for simulated process, and agent $i$ gets the reward equal to $\max_{a_i\in\AcSet_i} Q_i^{tr}(s_{i,t+1},a_i)$. One should choose the static planning horizon $k_{stat}$ do not exceed the visibility region.

To solve the simulated game, the RL agent solves model-based Q-learning by sampling transitions from the model $M_i$. When a sampled transition $(s_{i,t},a_{i,t},s_{i,t}',r_{i,t})$ is safe, the corresponding Q-function $Q_i^{(m)}(s_i,a_i)$ is updated according to \eqref{IQlearningUpdate}, {when the action $a_{i}$} is verified unsafe the shield $\phi_i$ replaces it with a backup policy $\pi_{backup}(o_i(s))$ and Q-function $Q(s_i, a_{i})$ is updated with a penalty \(r_{i,shield}\). For a more detailed description of the algorithm, let us refer the interested reader to our previous work \citep{jin2025predictive}.
Note that, since the simulated process is stationary, Q-learning will converge to an optimal Q-table $Q_i^{st}$ of the simulated process \citep{SuttonRLbook}. 

\subsection{Predictive shield for dynamic obstacles}

While $Q$-learning works well in stationary environments, in a non-stationary environment it may not converge. We hence propose to reformulate  \eqref{maincumrewardperagent}, as finite-horizon optimization problem, and use time-dependent $Q$-value functions in the $Q$-learning update rule.
Let us now choose the dynamic planning horizon $k_d$, such that $k_d\leq k_{stat}$. The cumulative reward \eqref{maincumrewardperagent} can be re-written as follows:
\begin{multline}
    \mathbb{E}_{\pi_i}\Big[\sum_{k=0}^{k_d} \gamma^k r_i(s_{i,t+k}, \phi_i( o_i(s_{t+k}), a_{i,t+k})) +\\
\gamma^{k_d+1}\max_{a_i\in \AcSet_i }Q_{\pi_i}(s_{i,t+k_d+1}, a_i)\Big].
\end{multline}
The last term in the equation above can be approximated with $Q^{st}(s_{i,t+k_d+1}, a_i)$:
\begin{multline}
\label{dynamicProblem}
\mathbb{E}_{\pi_i}\Big[\sum_{k=0}^{k_d} \gamma^k r_i(s_{i,t+k}, \phi_i( o_i(s_{t+k}), a_{i,t+k})) +\\ \gamma^{k_d+1}\max_{a_i\in \AcSet_i }\Qstat_i(s_{i,t+k_d+1}, a_i)\Big].
\end{multline}
The agent $i$ cannot predict the full state of the system $s_t,$ since it cannot predict the trajectories of the other agents. It is then searching for the policy $\pi_i$, maximizing the worst-case reward:
\begin{multline}
\label{worstcasedynamicProblem}
\mathbb{E}_{\pi_i}\big[\sum_{k=0}^{k_d}\gamma^k \min_{s_{t+k} \in \tilde R_{i,k}}r_i(s_{i,t+k}, \phi_i( o_i(s_{t+k}), a_{i,t+k})) +\\ \gamma^{k_d+1}\max_{a_i\in \AcSet_i }\Qstat_i(s_{i,t+k_d+1}, a_i)\big].
\end{multline}
where $\tilde R_{i,k}$ is defined as follows:
{\small\begin{equation}
\label{eq:worst_case_set}
\begin{split}
    \tilde R_{i,k} = \ & R_1(k|t, s_{1,t}) \times \dots \times R_{i-1}(k|t, s_{i-1,t}) \\
    & \times \{s_{i,t+k}\} \times R_{i+1}(k|t, s_{i+1,t}) \times \dots \times R_{N}(k|t, s_{N,t}).
\end{split}
\end{equation}}
Here $R_j(k|t, s_{i,t}), j \in \AgentSet $ are the reachable sets from $s_{i,t},$ defined at \eqref{reachSets}. Since we optimize over finite horizon $k_d$, in reality we only take into account the agents that are in $2k_d$ observability range. We then use Algorithm \ref{algo1} to find $\pi_i$ that maximizes \eqref{worstcasedynamicProblem}. 

\subsection{Model-Based Finite-Horizon Independent Q-Learning}
\label{ALGO}
\begin{algorithm}[t!]
\small
\caption{MB-FH-IQL}
\label{algo1}
\begin{algorithmic}[1]
\Require Pretrained Q-table $Q_i^{st}$, predictive horizon $k_d$, learning rate $\alpha$, discount $\gamma$, model $\mathcal{M}$, agent current state $s_c$, threshold $\varepsilon$

\State Initialize Q-tables $Q_i^0,\dots,Q_i^{k_d+1}$ with $Q_i^{st}$ \label{init}

\Repeat \label{repeat_start}
    \State $Q_i^{0prev} \gets Q_i^0$ 
    \State $s_{i,t} \gets s_c$ \label{start}
    
    \For{$k \in \{0,...,k_d\}$ }
        \State $a_{i,t+k} \gets \varepsilon$-greedy policy from $Q_i^k[s_{i,t+k}]$ \label{egreedy}
        \State $a_{i,safe} \gets a_{i,t+k}$ 
        
        \For{$s^\prime \in \tilde R_{i,k}$} 
            \If{$\phi_i(o_i(s^\prime), a_{i,t+k}) \neq a_{i,t+k}$} 
                \State $a_{i,safe} \gets \pi_{backup}(o_i(s^\prime))$ 
                \State \textbf{break} 
            \EndIf
        \EndFor
        
        \If{$\mathcal{M}(s_{i,t+k}, a_{i,safe})$ is not empty}
            \State $(s_{i,t+k+1}, r) \gets \mathcal{M}(s_{i,t+k}, a_{i,safe})$
        \Else
            \State \textbf{break}
        \EndIf
        
        \If{$a_{i,safe} = \pi_{backup}(o_i(s^\prime))$ }
            \State $r \gets r_{i,k,shield} (a_{i,t+k},s_{i,t+k})$
        \EndIf
        
        \Statex \hskip\algorithmicindent $\mathtt{UpdQ_k}(s_{i,t+k}, a_{i,t+k}, s_{i,t+k+1}, r, \alpha, \gamma)$
    \EndFor
    
\Until{$\|Q_i^0 - Q_i^{0prev}\|^2 < \varepsilon$}

\State \Return $Q_i^0$
\end{algorithmic}
\end{algorithm}

Algorithm~\ref{algo1} implements our \textit{Model-Based Finite-Horizon Independent Q-Learning} (MB-FH-IQL). Unlike standard IQL, the agent simulates its own evolution using an internal dynamics model $\mathcal{M}$ and assumes a \textit{worst-case evolution} of other agents \eqref{worstcasedynamicProblem}. 

Iteratively, the agent generates a $k_d$-step trajectory. Actions are selected via an $\varepsilon$-greedy policy and verified by a safety shield. If the shield overrides an action, the reward is replaced by the minimum possible reward in the reachable set $\tilde R_{i,k}$:
{\small\[
r_{i,k,shield} (a_{i,t+k},s_{i,t+k}) = \min_{s_k \in \tilde R_{i,k}}r_i(s_{i,t+k}, \phi_i( o_i(s_{t+k}), a_{i,t+k})).
\]}
The Q-values are then updated by propagating information forward along the horizon:
{\small\[
Q_k^{(m+1)}(s_k, a_k) = Q_k^{(m)} + \alpha \Big[ r_k + \gamma \max_{a'} Q_{k+1}^{(m)}(s_{k+1}, a') - Q_k^{(m)} \Big].
\]}
By augmenting the state space with the prediction step $k$, the value function incorporates non-stationary effects caused by dynamic obstacles.

The MB-FH-IQL algorithm is integrated into a full predictive shield architecture. The process first checks if the agent's current policy leads to static or dynamic safety violations within $k_{stat}$ or $k_d$ steps, respectively. If a violation is detected, infinite-horizon Q-learning is first employed on a local frame to converge to the Q-table  $Q_i^{st}$ with respect to static safety constraints only. Subsequently, MB-FH-IQL is executed only if the policy leads to dynamic safety violations within $k_d$ steps, using the Q-table that locally re-updated to avoid the static obstacles $Q_i^{st}$ as the terminal cost.
This architecture separates static and dynamic safety constraints for two main reasons.
First, to accelerate convergence regarding static constraints infinite horizon RL converges faster in stationary environments than finite horizon
Second, this separation optimizes computational costs by allowing distinct horizon lengths: while static constraints typically require longer horizons to anticipate structural deadlocks, dynamic collision avoidance is often effective with shorter horizons. Decoupling these processes avoids unnecessary computation.

\section{Communication-free conflict resolution protocol}

The MB-FH-IQL framework presented in Algorithm~\ref{algo1} seeks to reduce inter-agent interference through predictive shielding. However, due to decentralization, densely occupied or symmetric configurations may still generate livelocks. To mitigate these situations, we introduce a communication-free coordination protocol that probabilistically alternates between a nominal single-agent policy and a conservative multi-agent policy.

\begin{algorithm}[t!]
\caption{Conflict Resolution Protocol}
\small
\label{algo:prob_coord}
\begin{algorithmic}[1]

\Require Current state $s_{i,t}$, dynamic Q-table $Q_i^0$, static Q-table $Q_i^{\mathrm{st}}$, probability function $\varepsilon(\cdot)$, coordination count $c_i$
\Ensure Action to execute over two steps

\If{$\arg\max_{a^\prime} Q_i^{\mathrm{st}}(s_{i,t}, a^\prime)$ \text{is verified unsafe}} \label{alg2:Condition}

    \State $a_{i,\text{stat},1} \gets \phi_i(o_i(s_t), \argmax_{a'} Q_i^{\text{st}}(s_{i,t},a'))$ \label{alg2:single_agent1}
    \State \textbf{With probability $\varepsilon(c_i)$}:
    \State \quad Environment step with $a_i = a_{i,\text{stat},1}$  \label{alg2:single_aexec_a1}
    \State \quad $a_{i,\text{stat},2} \gets \phi_i(o_i(s_{t+1}), \argmax_{a'} Q_i^{\text{st}}(s_{i,t+1},a'))$ \label{alg2:single_agent2}
    \State \quad Environment step with $a_i = a_{i,\text{stat},2}$  \label{alg2:single_aexec_a2}

    \State \textbf{With probability $1-\varepsilon(c_i)$}:
    \State \quad $a_{i,1} \gets \text{$a_{bs}$}(o_i(s_{t}), Q_i^0,\{a_{i,\text{stat},1}\})$  \label{alg2:calc_a1}
    \State \quad Environment step with $a_i = a_{i,1}$   \label{alg2:aexec_a1}

    \State \quad $a_{i,2} \gets \text{$a_{bs}$}(o_i(s_{t+1}), Q_i^0,\mathcal{A}^{\text{ex}}_i(s_{i,t+1}, s_{i,t}))$\label{alg2:calc_a2}
    \State \quad Environment step with $a_i = a_{i,2}$ \label{alg2:exec_a2}

\EndIf

\end{algorithmic}
\end{algorithm}
The protocol is activated when the nominal action 
$a_{\text{nom}} = \arg\max_{a^\prime} Q_i^{\text{stat}}(s_{t}, a^\prime)$, computed from the static single-agent Q-table, is predicted to be unsafe (Line~\ref{alg2:Condition}). We define a switching probability $\varepsilon(c_i)$ as a logarithmically decaying function of the coordination count $c_i$, which measures how frequently an agent has been involved in coordination situations.This mechanism introduces differentiated behaviors among agents facing similar local configurations without requiring explicit communication or role assignment.

With probability $\varepsilon(c_i)$, the agent follows the nominal single-agent policy for two consecutive steps (Lines~\ref{alg2:single_agent1}--\ref{alg2:single_aexec_a2}). Otherwise, with probability $1-\varepsilon(c_i)$, the agent executes a conservative two-step coordination sequence (Lines~\ref{alg2:calc_a1}--\ref{alg2:exec_a2}). This branch relies on the constrained action-selection operator $a_{bs}(o_i(s), Q_i, \mathcal{A}^{\text{ex}}_i)$,
{\small\begin{equation}
\label{eq:best_safe_action}
a_{bs}(o_i(s), Q_i, \mathcal{A}^{\text{ex}}_i) =
\begin{cases}
    \underset{a_i \in \mathcal{A}_{i}^{s\setminus ex}}{\arg\max}\ Q_i(s_i, a_i) & \text{if } \mathcal{A}_{i}^{s\setminus ex} \neq \emptyset \\
    \pi_{backup}(o_i(s)) & \text{otherwise.}
\end{cases}
,
\end{equation}}
  where $\mathcal{A}_{i}^{s\setminus ex} = \{a_i\in\mathcal{A}_i\mid \, \varphi_i(o_i(s),a_i)=1 \ \land a_i \notin \mathcal{A}^{\text{ex}}_i\}$, i.e.  $\mathcal{A}^{\text{ex}}_i$ denotes the set of actions, we exclude from the safe actions. 
 
During the first step (Line~\ref{alg2:calc_a1}), the protocol excludes the nominal action selected in the alternate branch by setting $\mathcal{A}_{\text{ex}} = \{a_{i,stat,1}\}$. The agent is therefore encouraged to select an alternative safe action, which helps breaking the symmetry 

In the second step (Line~\ref{alg2:calc_a2}), the protocol excludes actions that would immediately return the agent to its previous state:
\begin{equation}
\label{eq:excluded_actions_dynamics}
\mathcal{A}^{\text{ex}}_i(s_i, s_{i,\text{prev}}) = \{ a_i \in \mathcal{A}_i \mid \mathcal{T}_i(s_i, a_i) = s_{i,\text{prev}} \}
\end{equation}

This constraint prevents immediate reversals and reduces short oscillatory behaviors. The resulting two-step sequence is inspired by the pseudo-goal coordination strategies introduced by \citep{jha2024spgp}. In particular, the conservative branch temporarily deviates from the goal, while the second-step exclusion constraint discourages an immediate return to the previous state. As a result, the agent is implicitly guided toward a temporary intermediate objective.
\section{Dec-POMDP Algorithm}

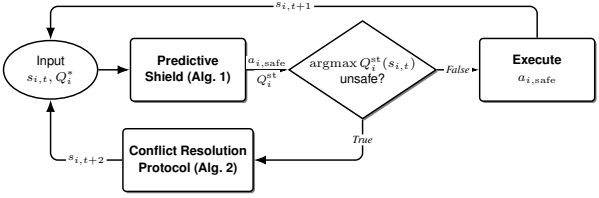
\begin{figure}[ht]
    \centering
    \resizebox{0.9\columnwidth}{!}{ 
    
    \begin{tikzpicture}[
        node distance=1.2cm and 0.8cm,
        font=\sffamily\small,
        >=Latex,
        process/.style={
            draw=black, thick, rectangle, rounded corners=2pt,
            minimum width=2.2cm, minimum height=1.2cm, 
            align=center, fill=white,
            drop shadow={opacity=0.15, shadow xshift=1pt, shadow yshift=-1pt},
            font=\sffamily\scriptsize
        },
        decision/.style={
            draw=black, thick, diamond, aspect=1.5,
            minimum width=2.5cm, minimum height=1cm,
            align=center, fill=white, inner sep=0pt,
            drop shadow={opacity=0.15, shadow xshift=1pt, shadow yshift=-1pt},
            font=\sffamily\scriptsize
        },
        input/.style={
            draw=black, thick, ellipse, 
            minimum width=1.8cm, minimum height=0.8cm,
            align=center, fill=white,
            font=\sffamily\scriptsize
        },
        lbl/.style={
            midway, fill=white, inner sep=1pt, font=\tiny\itshape, align=center
        }
    ]

        \node[input] (start) {Input\\$s_{i,t}, Q_i^*$};

        \node[process, right=0.6cm of start] (shield) {
            \textbf{Predictive}\\
            \textbf{Shield (Alg.~\ref{algo1})}
        };

        \node[decision, right=0.8cm of shield] (cond) {
            $\argmax Q_i^{\mathrm{st}}(s_{i,t})$\\
            unsafe?
        };

        \node[process, right=0.8cm of cond] (exec) {
            \textbf{Execute}\\
            $a_{i,\mathrm{safe}}$
        };

        \node[process, below=0.5cm of shield] (conflict) {
            \textbf{Conflict Resolution}\\
            \textbf{Protocol (Alg.~\ref{algo:prob_coord})}
        };

        \draw[->, thick] (start) -- (shield);

        \draw[->, thick] (shield) -- 
            node[lbl, above] {$a_{i,\mathrm{safe}}$} 
            node[lbl, below] {$Q_i^{\mathrm{st}}$}
            (cond);

        \draw[->, thick] (cond) -- node[lbl] {False} (exec);

        \draw[->, thick, rounded corners=4pt] (exec.north) 
            -- ++(0, 0.6) coordinate(topPath)
            -- node[lbl] {$s_{i,t+1}$} (topPath -| start.north) 
            -- (start.north);

        \draw[->, thick, rounded corners=4pt] (cond.south) 
            |- node[lbl, pos=0.25] {True} 
            (conflict.east);

        \draw[->, thick, rounded corners=4pt] (conflict.west) 
            -| node[lbl, pos=0.25] {$s_{i,t+2}$} 
            (start.south);

    \end{tikzpicture}
    } 
    \caption{Flowchart of algorithms and executions per time step for each agent $i$}
    \label{fig:horizontal_loop}
\end{figure}

Fig. \ref{fig:horizontal_loop} illustrates the proposed control architecture. The predictive shield operates as the primary control layer, filtering actions to mitigate dynamic interference while preserving goal-directed behavior. However, when the shield detects that the nominal policy is infeasible due to a crowded environment or a symmetric situation, the system transitions to the Conflict Resolution Protocol. To escape these local minima, the protocol employs a stochastic switching strategy where agents probabilistically alternate between a conservative multi-agent policy and the nominal trajectory. This induced stochasticity effectively breaks symmetry, minimizing livelocks. This architecture strictly guarantees that unsafe actions are never executed.

\section{Experiments}
\label{sec:experiments}
\begin{table*}[t!]
    \centering
    \caption{{Performance Comparison: IQL vs MIS vs Dyna Shield vs DMPS vs Ours.} ($H$: Horizon, $T_{\text{pre}}$: Pretrain Time, $T_{\text{dep}}$: Deployment Time)}
    \label{tab:results_comparison}
    
    \setlength{\tabcolsep}{2pt} 
    
    \resizebox{\textwidth}{!}{
    \begin{tabular}{l cccc ccccc ccccc ccccc ccccc}
        \toprule
        \multirow{2}{*}{\textbf{Exp.}} & 
        \multicolumn{4}{c}{\textbf{IQL}} & 
        \multicolumn{5}{c}{\textbf{MIS}} & 
        \multicolumn{5}{c}{\textbf{Dyna-Q Shield}} &  
        \multicolumn{5}{c}{\textbf{DMPS}} &         
        \multicolumn{5}{c}{\textbf{Ours}} \\
        
        \cmidrule(lr){2-5} \cmidrule(lr){6-10} \cmidrule(lr){11-15} \cmidrule(lr){16-20} \cmidrule(lr){21-25}
        
         & $T_{\text{pre}}(s)$ & $T_{\text{dep}}(s)$ & Ret. & Steps 
         & $k$ & $T_{\text{pre}}(s)$ & $T_{\text{dep}}(s)$ & Ret. & Steps 
         & $k$ & $T_{\text{pre}}(s)$ & $T_{\text{dep}}(s)$ & Ret. & Steps 
         & $k$ & $T_{\text{pre}}(s)$ & $T_{\text{dep}}(s)$ & Ret. & Steps 
         & $(k_{stat},k_{dyna})$ & $T_{\text{pre}}(s)$ & $T_{\text{dep}}(s)$ & Ret. & Steps \\
        \midrule
        
        \textbf{Exp 1} 
        & 531.1 & $1.91\times 10^{-5}$ & 180 & 11  
        & $1$ & 5.5 & $4.9\times10^{-4}$ & $-\infty$ & $\infty$ 
        & $7$ & 5.5& 0.27 & 151.2 & 26.8
        & $7$ & 5.5 & 0.1 & 180 & 11  
        & (7,4)& 5.5 & 0.05 & \textbf{180} & \textbf{11} \\ 
        
        \textbf{Exp 2} 
        & 1614.5 & $3.14\times 10^{-5}$& \textbf{360} & \textbf{11}
        & $1$ & 5.8 & $1.01\times10^{-4}$ & $-\infty$ & $\infty$
        & $7$ & 5.8 & 0.67 & 48.70 & 75.3
        & $7$ & 5.8 & 1.13 &  $-\infty$ & $\infty $
        & (7,2) & 5.8 & 0.41 & 328.5 & 31 \\ 
        
        \textbf{Exp 3} 
        & 3382.3 &$2.82 \times 10^{-3}$& -94 & $\infty$ 
        & $1$ & 6.95  & $2.04 \times 10^{-3}$& $-\infty$ & $\infty$ 
        & $7$ & 6.95  & 0.02 & -379.8& $\infty$ 
        & $7$ &6.95  & 0.88 & $-\infty$ & $\infty$ 
        & (7,2) & 6.95 & 2.6 & \textbf{268.4} & \textbf{105.2} \\
        \bottomrule
    \end{tabular}%
    }
\end{table*}
In this section, we evaluate the proposed method using the \texttt{gym-multigrid} framework~\citep{gym_multigrid}\footnote{https://github.com/ArnaudFickinger/gym-multigrid}. Our evaluation focuses on two distinct environments: a standard multi-agent path findingscenario and a constrained coin collection task, where agents must collect assigned coins while strictly avoiding those belonging to others. Across all experiments, the discrete action space is defined as $\mathcal{A} = \{ \mathtt{up}, \mathtt{down}, \mathtt{left}, \mathtt{right}\}$,and the backup policy is $\pi_{\text{backup}}(s) = \mathtt{stay}$, ensuring hard safety guarantees in non-adversarial environments where all agents are shielded, as agents can indefinitely hold their position without being collided with by other shielded agents. Finally, the reward function is defined as follows: $+100$ for reaching a goal or collecting a valid coin, $-10$ for inter-agent collisions, and $-100$ for hitting a static obstacle or collecting an invalid coin\textsuperscript{\monNumeroGitHub}.

\begin{figure}[htbp]
    \centering
    \begin{minipage}[b]{0.32\linewidth}
        \centering
        \includegraphics[width=\linewidth]{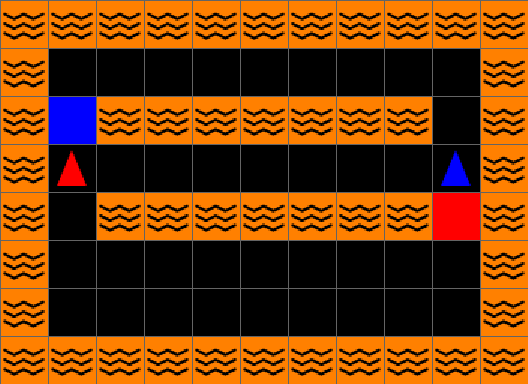}
        {\scriptsize (a) Exp1: Narrow corridor}
        \label{fig:narrow}
    \end{minipage}
    \hfill 
    \begin{minipage}[b]{0.32\linewidth}
        \centering
        \includegraphics[width=\linewidth]{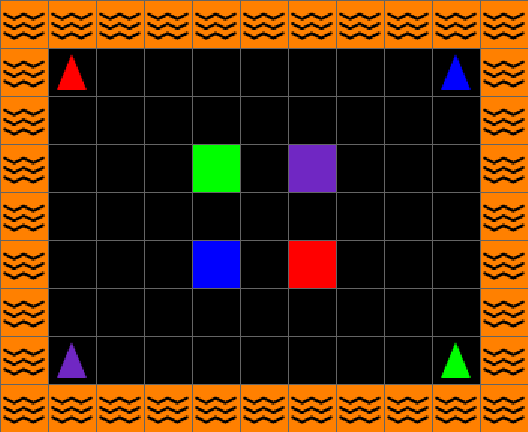}
        {\scriptsize (b) Exp2: Symmetric env.} 
        \label{fig:symmetric}
    \end{minipage}
    \hfill
    \begin{minipage}[b]{0.32\linewidth}
        \centering
        \includegraphics[width=\linewidth]{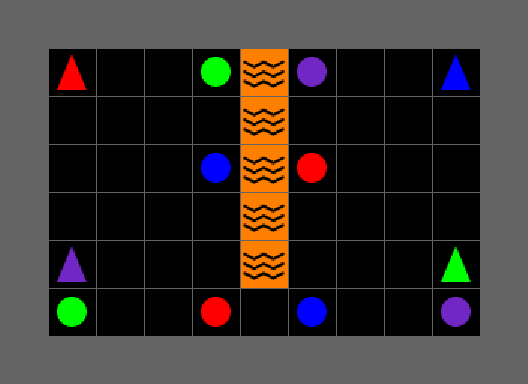}
        {\scriptsize (c) Exp3: Crowded passage}
        \label{fig:crowded}
    \end{minipage}
    
    \caption{In the two figures on the right, the agents are solving a path-finding game. The agents are represented by triangles, and their goals are shown as squares. In the third figure, the agents (triangles) need to collect the balls (circles) of the corresponding color.}
    \label{fig:canonical_env}
\end{figure}

We conduct two types of evaluation, first benchmarking our approach against four baselines on three canonical scenarios (Fig.~\ref{fig:canonical_env}). We include \textbf{IQL}~\citep{IQL} as an independent learning technique where agents are trained within the shared environment with collision penalties but without an explicit shield. The other three baselines are shielding techniques that utilize a Q-table $Q_i^{tr}$ pre-trained in a single-agent, obstacle-free environment; for these methods, the agent initial positions and objectives remain identical between the pre-training and deployment phases. Within this group, the \textbf{Minimal Interference Shield (MIS)}~\citep{ElSayedAly} is a reactive method that follows the $Q_i^{tr}$ policy and only invokes $\pi_{\text{backup}}$ upon imminent violation. In contrast, the remaining two are predictive techniques: the local planner from \textbf{Dynamic Model Predictive Shielding (DMPS)}~\citep{BanerjeeDynamicMPSShield} is a finite-horizon planner that treats static and dynamic constraints identically, while the \textbf{Predictive Safety Shield (Dyna-Q Shield)}~\citep{jin2025predictive} treats other agents as static obstacles during its updates. Finally, for all shielding-based methods, we enforce a decentralized safety guarantee that prohibits entry into any cell reachable by a neighbor in a single step to ensure hard safety against dynamic constraints.

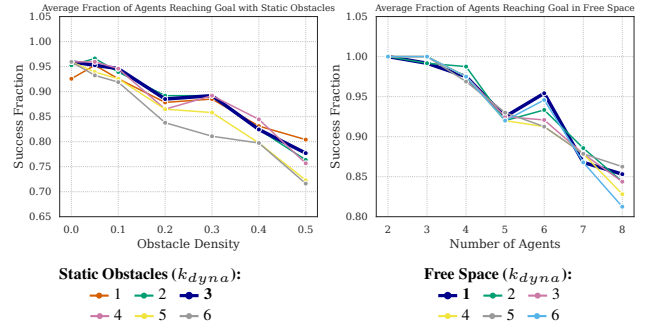
\begin{figure}[!t]
    \centering
    \includegraphics[width=0.95\columnwidth]{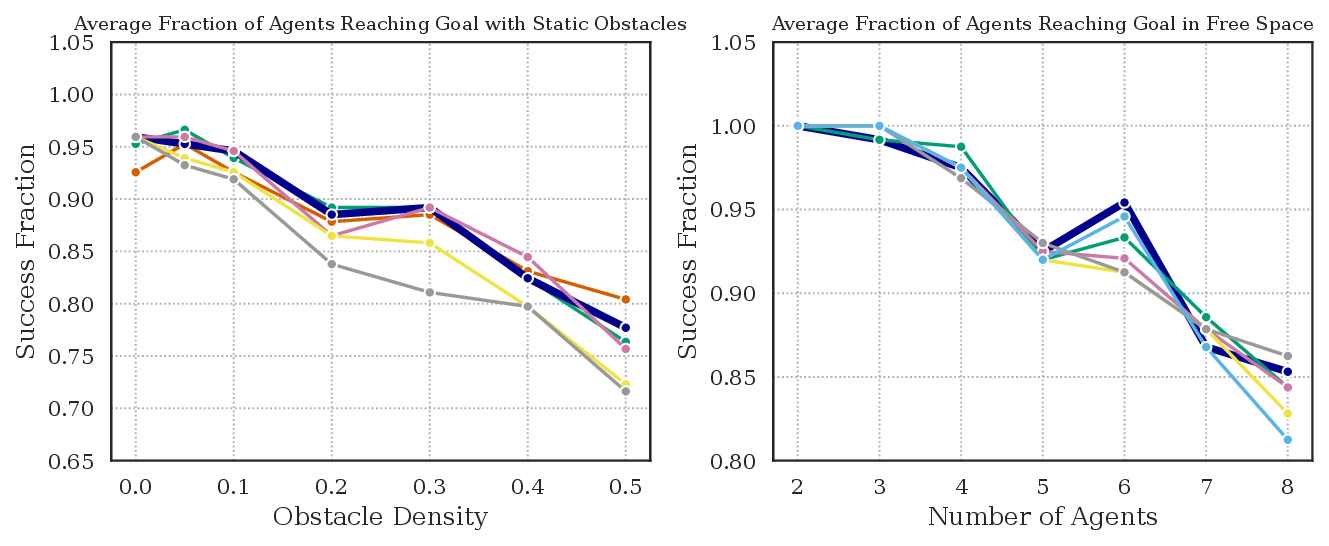}
    
    \vspace{0.1cm} 
    
    \begin{minipage}{0.48\columnwidth}
        \centering
        \tiny 
        \textbf{Static Obstacles ($k_{dyna}$):}\\
        \tikz[baseline=-0.5ex]{\draw[kOne, thick] (0,0) -- (0.25,0) node[midway, circle, fill, inner sep=0.8pt]{};} 1 \hspace{0.05cm}
        \tikz[baseline=-0.5ex]{\draw[kGreen, thick] (0,0) -- (0.25,0) node[midway, circle, fill, inner sep=0.8pt]{};} 2 \hspace{0.05cm}
        \tikz[baseline=-0.5ex]{\draw[kBest, line width=1.2pt] (0,0) -- (0.25,0) node[midway, circle, fill, inner sep=1pt]{};} \textbf{3}\\
        \tikz[baseline=-0.5ex]{\draw[kPink, thick] (0,0) -- (0.25,0) node[midway, circle, fill, inner sep=0.8pt]{};} 4 \hspace{0.05cm}
        \tikz[baseline=-0.5ex]{\draw[kYellow, thick] (0,0) -- (0.25,0) node[midway, circle, fill, inner sep=0.8pt]{};} 5 \hspace{0.05cm}
        \tikz[baseline=-0.5ex]{\draw[kGrey, thick] (0,0) -- (0.25,0) node[midway, circle, fill, inner sep=0.8pt]{};} 6
    \end{minipage}
    \hfill
    \begin{minipage}{0.48\columnwidth}
        \centering
        \tiny 
        \textbf{Free Space ($k_{dyna}$):}\\
        \tikz[baseline=-0.5ex]{\draw[kBest, line width=1.2pt] (0,0) -- (0.25,0) node[midway, circle, fill, inner sep=1pt]{};} \textbf{1} \hspace{0.05cm}
        \tikz[baseline=-0.5ex]{\draw[kGreen, thick] (0,0) -- (0.25,0) node[midway, circle, fill, inner sep=0.8pt]{};} 2 \hspace{0.05cm}
        \tikz[baseline=-0.5ex]{\draw[kPink, thick] (0,0) -- (0.25,0) node[midway, circle, fill, inner sep=0.8pt]{};} 3\\
        \tikz[baseline=-0.5ex]{\draw[kYellow, thick] (0,0) -- (0.25,0) node[midway, circle, fill, inner sep=0.8pt]{};} 4 \hspace{0.05cm}
        \tikz[baseline=-0.5ex]{\draw[kGrey, thick] (0,0) -- (0.25,0) node[midway, circle, fill, inner sep=0.8pt]{};} 5 \hspace{0.05cm}
        \tikz[baseline=-0.5ex]{\draw[kCyan, thick] (0,0) -- (0.25,0) node[midway, circle, fill, inner sep=0.8pt]{};} 6
    \end{minipage}
    
    \vspace{0.1cm}
    
    \caption{ Average fraction of agents reaching their goal under static obstacles (4 agents, \(k_{stat}=7\)) and in free space environments for different horizons}
    \label{fig:success_fraction}
\end{figure}

Second a scalability analysis on the multi-agent path finding task within a $10\times10$ grid through two parametric tests where each configuration is evaluated over 45 random initialization: the first varies the number of agents in free, while the second fixes the number of agents to four and varies the static obstacle density with $k_{\text{stat}}=7$. As shown in Figure~\ref{fig:success_fraction}, an increase in the density of either agents or static obstacles generally leads to a lower fraction of agents reaching their goals. To identify the most robust horizon across these varying conditions, we introduce a cumulative regret metric $R(k_{\text{dyna}})$, defined as:
\begin{equation}
R(k_{\text{dyna}}) = \sum_{c \in \mathcal{C}} \left( \max_{k} S(k, c) - S(k_{\text{dyna}}, c) \right)
\end{equation}
where $\mathcal{C}$ represents the set of test configurations and $S(k, c)$ denotes the success rate. In environments without static obstacles (Fig.~\ref{fig:success_fraction}), we observe that all values of $k_{\text{dyna}}$ perform quite similarly, with $k_{\text{dyna}}=1$ maintaining a slight edge while being preferable from a computational complexity perspective. However, the introduction of static obstacles (Fig.~\ref{fig:success_fraction}) shifts this trend, making a mid-length horizon of $k_{\text{dyna}}=3$ the most effective choice. Since $k_{\text{stat}}$ is already set to a long horizon to manage navigation around walls, the primary role of $k_{\text{dyna}}$ in these settings is to anticipate the presence of other agents before committing to narrow passages or bottlenecks. Without this look-ahead, agents may enter restricted spaces simultaneously, leading to situations that are difficult to resolve. These results illustrate a trade-off between look-ahead sufficiency and over-conservatism. While a mid-length horizon helps agents coordinate before entering constrained areas, excessively long horizons can lead to performance drops in dense settings. We attribute this to the accumulation of uncertainty over time, which causes the shielding mechanism to become overly sensitive, eventually inducing freezing behaviors or forcing agents to deviate excessively from their paths.

\section{Conclusion}

This paper proposed a fully decentralized framework for the safe deployment of independently trained RL agents in shared environments. By integrating a predictive safety shield and a communication-free conflict resolution protocol, our approach enables agents to adapt their policies online while guaranteeing safety and minimize the number of livelocks. Experimental results show that the proposed method outperforms shielding baselines in terms of task completion and independent learning approaches in terms of computation time.

Nevertheless, several challenges remain. First, the prediction horizon depends on the environment and must currently be chosen before deployment; future work should investigate adaptive horizon selection during execution. Second, the worst-case estimation makes the approach conservative, even with reward shaping, motivating the exploration of alternative methods such as trajectory estimation using conformal prediction \citep{lindemann2025formal}. Finally, the impact of model mismatch between training and deployment environments remains unclear and requires further investigation.

\section*{Acknowledgments}
This work was supported by the French National Research Agency (ANR) under grant ANR-22-EXES-0013. The authors also acknowledge Ferdinand Plesse-Costa for his assistance.
\bibliography{ifacconf}

\begin{thebibliography}{22}
\providecommand{\natexlab}[1]{#1}
\providecommand{\url}[1]{\texttt{#1}}
\providecommand{\urlprefix}{URL }
\expandafter\ifx\csname urlstyle\endcsname\relax
  \providecommand{\doi}[1]{doi:\discretionary{}{}{}#1}\else
  \providecommand{\doi}{doi:\discretionary{}{}{}\begingroup \urlstyle{rm}\Url}\fi

\bibitem[{Banerjee et~al.(2024)Banerjee, Rahmani, Biswas, and Dillig}]{BanerjeeDynamicMPSShield}
Banerjee, A., Rahmani, K., Biswas, J., and Dillig, I. (2024).
\newblock Dynamic model predictive shielding for provably safe reinforcement learning.
\newblock In \emph{Proceedings of the 38th International Conference on Neural Information Processing Systems}, NIPS '24.

\bibitem[{Bernstein et~al.(2000)Bernstein, Zilberstein, and Immerman}]{BernsteinPOMDP}
Bernstein, D.S., Zilberstein, S., and Immerman, N. (2000).
\newblock The complexity of decentralized control of markov decision processes.
\newblock In \emph{Proceedings of the Sixteenth Conference on Uncertainty in Artificial Intelligence}, UAI'00, 32--37.

\bibitem[{Brorholt et~al.(2025)Brorholt, Larsen, and Schilling}]{BrorholtCompositionalShielding}
Brorholt, A.H., Larsen, K.G., and Schilling, C. (2025).
\newblock Compositional shielding and reinforcement learning for multi-agent systems.
\newblock In \emph{Proceedings of the 24th International Conference on Autonomous Agents and Multiagent Systems}, AAMAS '25, 399--407.

\bibitem[{Carr et~al.(2025)Carr, Bakirtzis, and Topcu}]{Bakirtzis}
Carr, S., Bakirtzis, G., and Topcu, U. (2025).
\newblock Compositional shield synthesis for safe reinforcement learning in partial observability.
\newblock \emph{IEEE Open Journal of Control Systems}, 4, 373--384.

\bibitem[{Chandra et~al.(2025)Chandra, Zinage, Bakolas, and et~al.}]{ma2024deadlock}
Chandra, R., Zinage, V., Bakolas, E., and et~al. (2025).
\newblock Deadlock-free, safe, and decentralized multi-robot navigation in social mini-games via discrete-time control barrier functions.
\newblock \emph{Autonomous Robots}, 49(1), 12.

\bibitem[{ElSayed-Aly et~al.(2021)ElSayed-Aly, Bharadwaj, Amato, Ehlers, Topcu, and Feng}]{ElSayedAly}
ElSayed-Aly, I., Bharadwaj, S., Amato, C., Ehlers, R., Topcu, U., and Feng, L. (2021).
\newblock Safe multi-agent reinforcement learning via shielding.
\newblock In \emph{Proceedings of the 20th International Conference on Autonomous Agents and MultiAgent Systems}, AAMAS '21, 483--491.

\bibitem[{Fickinger(2020)}]{gym_multigrid}
Fickinger, A. (2020).
\newblock Multi-agent gridworld environment for openai gym.

\bibitem[{Garg et~al.(2024)Garg, Zhang, So, Dawson, and Fan}]{ChuChuFAn}
Garg, K., Zhang, S., So, O., Dawson, C., and Fan, C. (2024).
\newblock Learning safe control for multi-robot systems: Methods, verification, and open challenges.
\newblock \emph{Annual Reviews in Control}, 57, 100948.

\bibitem[{Grover et~al.(2023)Grover, Liu, and Sycara}]{grover2023deadlock}
Grover, J., Liu, C., and Sycara, K. (2023).
\newblock Deadlock analysis and resolution in multi-robot systems.
\newblock \emph{The International Journal of Robotics Research}, 42(1-2), 57--87.

\bibitem[{Guo and Dimarogonas(2015)}]{GuoMeng}
Guo, M. and Dimarogonas, D.V. (2015).
\newblock Multi-agent plan reconfiguration under local ltl specifications.
\newblock \emph{Int. J. Rob. Res.}, 34(2), 218--235.

\bibitem[{Jha et~al.(2024)Jha, Gupta, Rawat, and Kumar}]{jha2024spgp}
Jha, A., Gupta, T., Rawat, S., and Kumar, G. (2024).
\newblock Strategic pseudo-goal perturbation for deadlock-free multi-agent navigation in social mini-games.
\newblock 264--269.

\bibitem[{Jin et~al.(2025{\natexlab{a}})Jin, Krasowski, and Vanneaux}]{jin2025predictive}
Jin, P., Krasowski, H., and Vanneaux, E. (2025{\natexlab{a}}).
\newblock Predictive safety shield for dyna-q reinforcement learning.
\newblock In \emph{2025 European Control Conference (ECC)}, 2173--2179. IEEE.

\bibitem[{Jin et~al.(2025{\natexlab{b}})Jin, Chen, Lin, Song, and Wierman}]{pmlr-v258-jin25a}
Jin, R., Chen, Z., Lin, Y., Song, J., and Wierman, A. (2025{\natexlab{b}}).
\newblock Approximate global convergence of independent learning in multi-agent systems.
\newblock In \emph{Proceedings of The 28th International Conference on Artificial Intelligence and Statistics}, 2818--2826.

\bibitem[{Krasowski et~al.(2023)Krasowski, Thumm, M{\"u}ller, Sch{\"a}fer, Wang, and Althoff}]{krasowski2023provably}
Krasowski, H., Thumm, J., M{\"u}ller, M., Sch{\"a}fer, L., Wang, X., and Althoff, M. (2023).
\newblock Provably safe reinforcement learning: Conceptual analysis, survey, and benchmarking.
\newblock \emph{Transactions on Machine Learning Research}.

\bibitem[{Li and Bastani(2020)}]{Bastani}
Li, S. and Bastani, O. (2020).
\newblock Robust model predictive shielding for safe reinforcement learning with stochastic dynamics.
\newblock In \emph{2020 IEEE International Conference on Robotics and Automation (ICRA)}, 7166--7172.

\bibitem[{Lindemann et~al.(2025)Lindemann, Zhao, Yu, Pappas, and Deshmukh}]{lindemann2025formal}
Lindemann, L., Zhao, Y., Yu, X., Pappas, G.J., and Deshmukh, J.V. (2025).
\newblock Formal verification and control with conformal prediction: Practical safety guarantees for autonomous systems.
\newblock \emph{IEEE Control Systems}, 45(6), 72--122.
\newblock \doi{10.1109/MCS.2025.3611545}.

\bibitem[{Oroojlooy and Hajinezhad(2022)}]{Oroojlooy2022}
Oroojlooy, A. and Hajinezhad, D. (2022).
\newblock A review of cooperative multi-agent deep reinforcement learning.
\newblock \emph{Applied Intelligence}, 53(11), 13677--13722.

\bibitem[{Sheng et~al.(2024)Sheng, Parker, and Feng}]{SafePOMDPOnlinePlanningviaShielding}
Sheng, S., Parker, D., and Feng, L. (2024).
\newblock Safe pomdp online planning via shielding.
\newblock In \emph{2024 IEEE International Conference on Robotics and Automation (ICRA)}, 126--132.

\bibitem[{Sutton and Barto(2018)}]{SuttonRLbook}
Sutton, R.S. and Barto, A.G. (2018).
\newblock \emph{Reinforcement Learning: An Introduction}.
\newblock A Bradford Book, Cambridge, MA, USA.

\bibitem[{Tan(1993)}]{IQL}
Tan, M. (1993).
\newblock Multi-agent reinforcement learning: independent versus cooperative agents.
\newblock In \emph{Proceedings of the Tenth International Conference on International Conference on Machine Learning}, ICML'93, 330--337.

\bibitem[{Tang et~al.(2025)Tang, Abbatematteo, Hu, Chandra, Martín-Martín, and Stone}]{Tang2025_DeepRL_Robotics}
Tang, C., Abbatematteo, B., Hu, J., Chandra, R., Martín-Martín, R., and Stone, P. (2025).
\newblock Deep reinforcement learning for robotics: A survey of real-world successes.
\newblock \emph{Annual Review of Control, Robotics, and Autonomous Systems}, 8(1), 153--188.

\bibitem[{Xiao et~al.(2023)Xiao, Lyu, and Dolan}]{XiaoModelBasedDynamicShielding}
Xiao, W., Lyu, Y., and Dolan, J. (2023).
\newblock Model-based dynamic shielding for safe and efficient multi-agent reinforcement learning.
\newblock In \emph{Proceedings of the 2023 International Conference on Autonomous Agents and Multiagent Systems}, AAMAS '23, 1587--1596.

\end{thebibliography}

\theendnotes
\end{document}